\documentclass[aps,prl,twocolumn,superscriptaddress]{revtex4-2}
\usepackage{mathtools}
\usepackage{xcolor}
\usepackage{graphicx}
\usepackage{amsmath}
\usepackage{amssymb}

\newcommand{\av}[1]{\left\langle #1 \right\rangle}              
\newcommand{\e}{\varepsilon}              %
\newcommand{\vp}{{\varphi}}              %
\newcommand{\w}{{\omega}}              %
\newcommand{\W}{{\Omega}}              %

\begin{document}

\title{Coherence properties of collective modes in ensembles of oscillators}

 
\author{Arkady Pikovsky}
 \email{pikovsky@uni-potsdam.de}
 \affiliation{Department of Physics and Astronomy, University of Potsdam, Karl-Liebknecht-Str. 24/25, 14476, Potsdam-Golm, Germany}
 \author{Franco Bagnoli}
 \email{franco.bagnoli@unifi.it}
 \affiliation{Department of Physics and Astronomy and CSDC, University of Florence, via G. Sansone 1, I-50019, Sesto Fiorentino, Italy}
 \email{franco.bagnoli@unifi.it}
 \affiliation{Istituto Nazionale di Fisica Nucleare, Sezione di Firenze, via G. Sansone 1, I-50019, Sesto Fiorentino, Italy}
 \author{Stefano Iubini}
 \email{stefano.iubini@cnr.it}
 \affiliation{Istituto dei Sistemi Complessi, Consiglio Nazionale delle Ricerche, via Madonna del Piano 10, I-50019, Sesto Fiorentino, Italy} 
  \affiliation{Istituto Nazionale di Fisica Nucleare, Sezione di Firenze, via G. Sansone 1, I-50019, Sesto Fiorentino, Italy}

\date{\today}

\begin{abstract}
Synchronization transition in oscillatory networks manifests itself as the appearance of a periodic global mode. While perfect in the thermodynamic limit, this mode fluctuates for finite ensembles. We characterize the coherence of this mode in terms of the phase diffusion constant. In several examples, we always observed normal diffusion, but the dependence of the diffusion constant on the system size $D\sim N^{-\mu}$ depends on the nature of coupled units: for coupled chaotic systems  $\mu=1$, while for coupled periodic oscillators we observe, depending on the particular model, $\mu=2$ and $\mu=2.5$. These large values of the power index are attributed to the size-dependence of collective chaos in the finite ensemble, which disappears in the thermodynamic limit. We also show that in the standard Kuramoto model for a symmetric set of frequencies, there is an additional transition to a symmetric chaotic state with vanishing diffusion of the global phase.
\end{abstract}

\maketitle

\textit{Introduction.}  
Clocks are not perfect: due to ambient noise, the coherence time is finite. The phase of the periodic self-sustained oscillators diffuses due to noise, and the diffusion constant, which has the dimension of inverse time, determines the coherence time and the width of the spectral peak~\cite{Bershtein-38,rubiola2008phase}. This concept can also be applied to collective oscillating modes appearing due to synchronization of individual units in large ensembles~\cite{Pikovsky-Rosenblum-Kurths-01,Acebron-etal-05}. Such transitions are usually theoretically treated in the thermodynamic limit of an infinite number of units~\cite{Kuramoto-75,Kuramoto-84}. In this limit, the transition to synchrony appears as a transition from a steady to oscillating mean field (typically via a Hopf bifurcation~\cite{Kuramoto-75,Ott-Antonsen-08,Dietert-16}). A periodic regime in the thermodynamic limit has infinite coherence time (except for cases where the collective mode in the thermodynamic limit is chaotic~\cite{Nakagawa-Kuramoto-94,chabanol1997collective,clusella2019between,clusella2020irregular}; we do not consider such regimes in this Letter).

In experimental realizations of the synchronization transition, the number of interacting units $N$ is necessarily finite, and in many cases rather small: 64 chemical oscillators in \cite{kiss2002emerging,zhai2005desynchronization};  72 electronic oscillators in \cite{temirbayev2013autonomous}; 108 and 690 Josephson junctions in~\cite{Barbara_etal-99}; 40 photochemical oscillators in \cite{taylor2008clusters}; 1600 chemical Belousov-Zhabotinsky oscillators in~\cite{totz2018spiral}; several hundreds of lasers in \cite{mahler2020experimental}; 30 metronoms in \cite{martens2013chimera}. In some cases, the number of interacting units is itself a bifurcation parameter. For example, the collective mode of the Millennium Bridge in London appeared when the number of walkers exceeded $N=164$~\cite{dallard2001london}.

For a finite number of coupled elements, the appearing global mode is subject to finite-size fluctuations (unless the full locking at a large coupling strength occurs). These fluctuations have been explored numerically and theoretically in ~\cite{daido1987scaling,daido1990intrinsic,%
Pikovsky-Ruffo-99,hong2005collective,hong2007entrainment,hildebrand2007kinetic,buice2007correlations,%
tang2011synchronize,Nishikawa_etal-14,Hong_etal-15,gottwald2017finite,Peter-Pikovsky-18,suman2024finite,omel2025mean}. However, in most cases, the focus was not on the coherence properties of the phase of the collective mode, but on the finite-size effects in the amplitude dynamics. Only in Refs.~\cite{Pikovsky-Ruffo-99,gottwald2017finite} has the phase diffusion constant been demonstrated to scale $\sim N^{-1}$ with the population size, for coupled noisy oscillators. Another situation in which the diffusion of the phase of the collective mode was studied is the Hamiltonian mean-field model~\cite{Ginelli_etal-11,filho2020classical}. There, due to momentum conservation, the dominant mode of motion of the collective phase is a ballistic one, with a transition to normal diffusion at a time that diverges with the system size.  

The goal of this Letter is to explore properties of the diffusion of the collective mode in ensembles of deterministic units. We demonstrate that coupled chaotic oscillators produce a regular mode with phase diffusion scaling $\sim N^{-1}$, like for coupled noisy units~\cite{Pikovsky-Ruffo-99,gottwald2017finite}. On the contrary, for ensembles of periodic oscillators, we observe much stronger decay of the diffusion with the system size. We also describe a remarkable transition, due to symmetry, in the standard Kuramoto system to a fully phase-coherent (but still chaotic) regime with quenched diffusion.

Before proceeding to the consideration of particular systems, let us discuss the general conditions for the existence of the global phase. Suppose that in the thermodynamic limit, there is a Hopf-type transition from static to oscillating mean field at critical coupling $\e_c$. One usually characterizes the transition with the amplitude of global oscillations $R$, which is zero for $\e<\e_c$ and is positive for $\e>\e_c$.  Fluctuations due to finite-size effects smear the transition, so that the amplitude $R$ fluctuates and is non-zero at all couplings, and the average of $R$ grows rapidly around $\e_c$. Still, as long as the amplitude of collective oscillations can vanish (or be very small) at some instants of time, the phase of these oscillations is ill-defined. Only when the amplitude $R$ is distinct from zero is the phase well-defined. The criterion for the onset of the global oscillating mode based on finding the coupling strength $\e_R$ beyond which $R_{min}>\gamma$ for some small threshold $\gamma$, has been suggested and used in experiments \cite{Temirbayev_etal-12,temirbayev2013autonomous} and numerically explored in \cite{peter2018transition}. Clearly, the threshold for existence of the global collective mode $\e_R$ depends on the system size and $\e_R$ becomes closer to $\e_c$ from above as $N\to\infty$;  this threshold defines the domain with a well-defined phase where the phase diffusion properties can be explored.  

\textit{Kuramoto-Sakaguchi ensemble.}
Our first model is the Kuramoto-Sakaguchi (KS) system of globally coupled phase oscillators~\cite{Sakaguchi-Kuramoto-86}
\begin{equation}
\dot\vp_k=\w_k+\frac{\e}{N}\sum_{j=1}^N \sin(\vp_j-\vp_k-\alpha),\;k=1,\ldots N\;.
\label{eq:ks}
\end{equation}
The Kuramoto order parameter
\begin{equation}
Z(t)=N^{-1}\sum_k\exp[i\vp_k(t)]=R(t)\exp[i\Theta(t)]
\label{eq:ksop}
\end{equation}
can be considered as the complex amplitude of the collective mode, where $R$ is the real amplitude and $\Theta$ is the phase of the collective global mode. We expect that for $\e>\e_R$, where the amplitude $R$ is separated from zero, the phase $\Theta$ is well-defined, and one can calculate its averaged rotation velocity $\Omega=\lim_{T\to\infty} T^{-1}(\Theta(T)-\Theta(0))$, as well as the diffusion constant $D=\lim_{T\to\infty} T^{-1}\langle (\Theta(T)-\Theta(0)-\Omega T)^2 \rangle$. The relevant parameter in problem \eqref{eq:ks} is the distribution of frequencies $g(\omega)$. In simple cases like Gaussian and Cauchy distributions (because of rotational invariance, one can set the center of the distribution at $\omega=0$), transition to synchrony in the thermodynamic limit occurs via a Hopf bifurcation to a state with $R=const$ and $\Omega=const$. 

\begin{figure}
\centering
\includegraphics[width=\columnwidth]{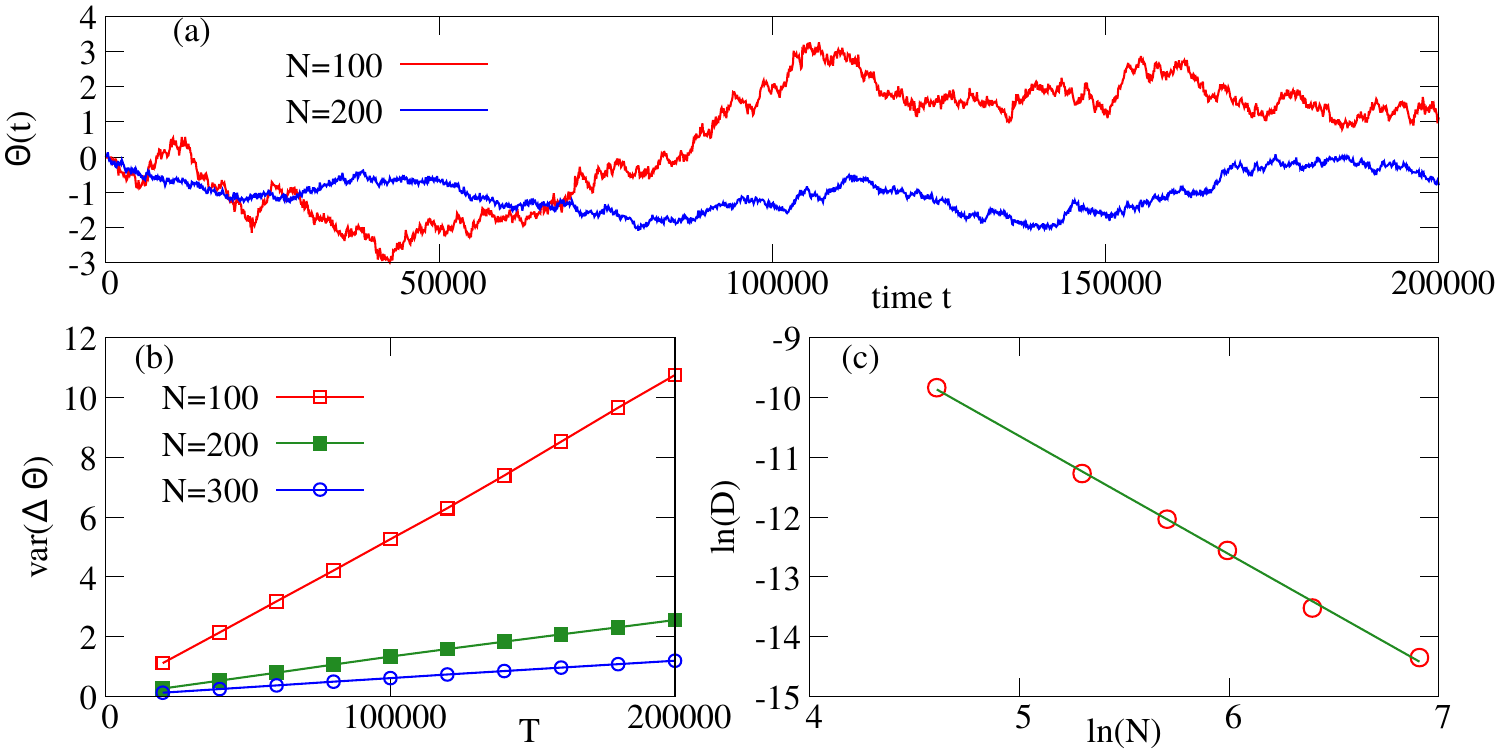}
\caption{Panel (a): Examples of phase variations $\Theta(t)-\Omega t$ for $N=100,\;200$. Panel (b): examples of dependencies of the phase variance $\langle (\Theta(T)-\Theta(0)-\Omega T)^2 \rangle$ vs time lag $T$; these dependencies are linear with high accuracy, indicating normal diffusion of the phase. Panel (c): log-log plot of the diffusion constants $D$ vs $N$ (in the range $100\leq N\leq 1000$). The best fit (green line) has slope $1.98$.}
\label{fig:1}
\end{figure}

In the simulation of a finite ensemble, one has to fix a way of choosing frequencies $\w_k$. We first consider a regular sampling~\cite{daido1987scaling,hong2007entrainment,Nishikawa_etal-14,Peter-Pikovsky-18,omel2025mean} according to relation $\int_{-\infty}^{\omega_k} g(\xi)d\xi=(k-1/2)/N$. This yields just one system for each $N$. Below we adopt for $g(\w)$ a Gaussian distribution with unit variance. Fig.~\ref{fig:1} shows an example of the phase diffusion and the dependence of the diffusion constant $D$ on the ensemble size $N$, for parameters $\alpha=\pi/4$, $\e=2.5$. The results show that with good accuracy, the law $D\sim N^{-2}$ holds. This means that the diffusion constant decreases much more strongly with the system size than the variance of amplitude variations, which decreases
in the supercritical regime as $\sim N^{-1}$~\cite{daido1987scaling,Hong_etal-15}. We attribute this strong drop of the diffusion constant to weak chaos, to be discussed below.

\begin{figure}
\centering
\includegraphics[width=\columnwidth]{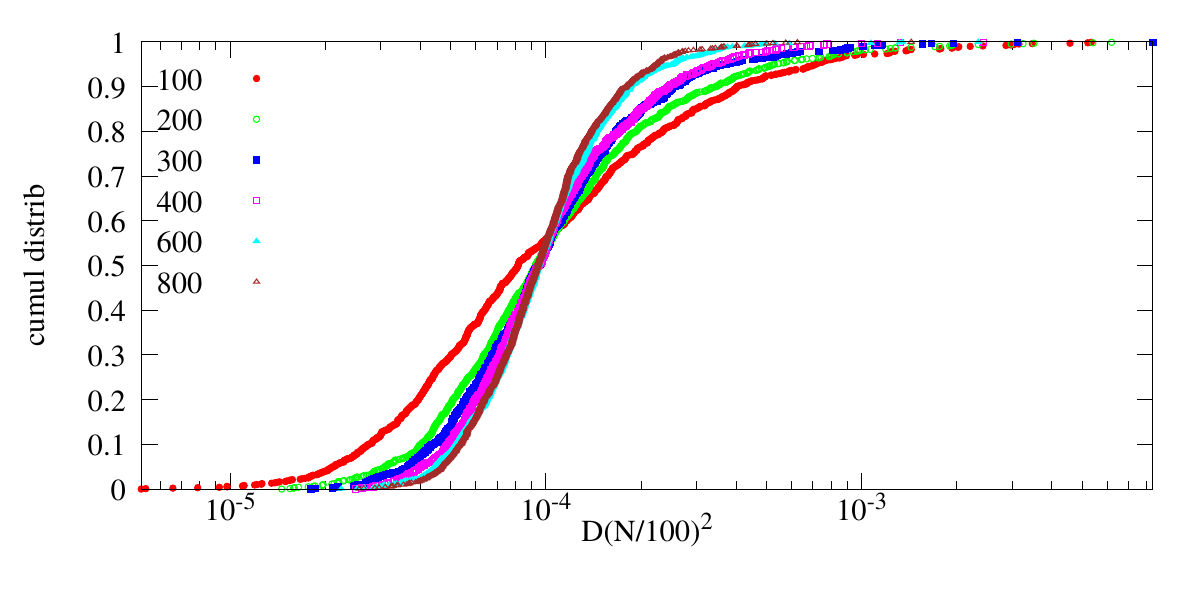}
\caption{Cumulative distributions of the diffusion constant $D$ for different system sizes, plotted for rescaled quantity $D (N/100)^2$. Parameters are the same as in the case of regular sampling: $\e=2.5$, $\alpha=\pi/4$.}
\label{fig:2}
\end{figure}

Above, we considered regular sampling, so that, for each $N$, there is a unique system. If, in contrast, one uses random sampling, then for each $N$ there is an ensemble of different systems with different values of the diffusion constant. To compare these constants at different $N$, we calculated the distribution functions of $D$ at different $N$, using $\approx 1000$ random samples for each $N$. Because the spread of diffusion constant at given $N$ is large, we compare in Fig.~\ref{fig:2} the distributions of $\log D$. To check for the scaling $\sim N^{-2}$, we depict the cumulative distributions of the rescaled diffusion constant $D (N/100)^2$ (so that for the smallest size $N=100$ there is no rescaling).  One can see that the plots intersect nearly at the same point at the level $1/2$, thus confirming the scaling $D\sim N^{-2}$.

\begin{figure}
\centering
\includegraphics[width=0.8\columnwidth]{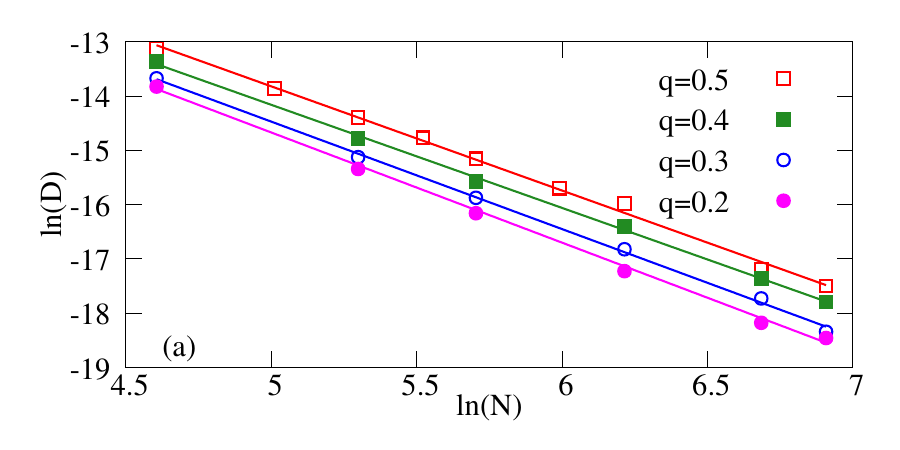}
\includegraphics[width=0.8\columnwidth]{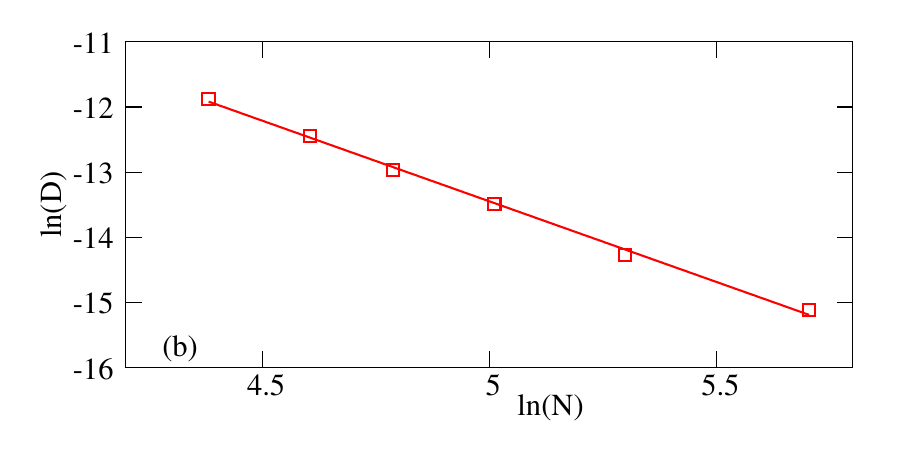}
\includegraphics[width=0.8\columnwidth]{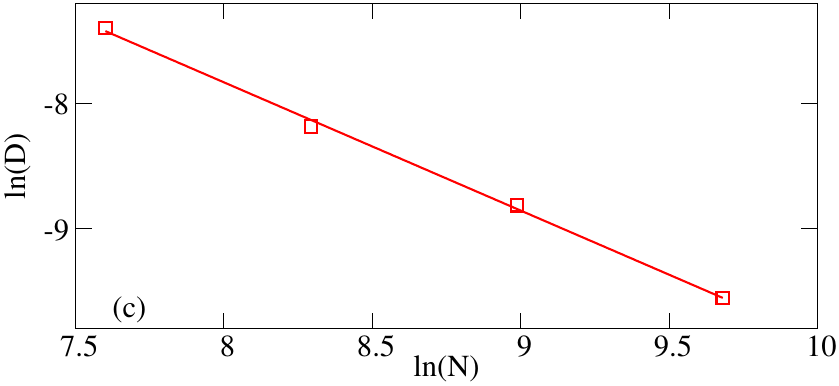}
\caption{System size dependencies (log-log plots) of the global phase diffusion constant for three models discussed in the text and formulated in the Appendix. Panel (a): nonlinearly coupled SL oscillators~\eqref{eq:sl1}, for four values of the frequency distribution half-width $q$. The best fit slopes are (from top to bottom) $1.92,\;1.9,\;1.98,\;2.03$. Panel (b): coupled ML system of neural spiking~\eqref{eq:mldet2}. The slope of the best fit is $2.48$. Panel (c): coupled chaotic Roessler systems~\eqref{eq:roes}. The best fit slope is $1.02$.}
\label{fig:3}
\end{figure}

\textit{Other examples of collective modes in ensembles of chaotic and regular units.}
The KS is a representative example of a synchronization transition, but one cannot expect its quantitative properties to be universally valid. Here we present evaluations of the coherence of the global mode for three other models (their equations are given in the Appendix). The first model that exhibits qualitatively similar synchronization transition is the Morris-Lecar (ML) neuron model~\eqref{eq:mldet2}. A single neuron is in spiking oscillatory states, and upon a strong enough coupling, a global periodic field appears in an ensemble. Some neurons with natural frequencies close to that of the mean field are locked, others are not entrained; this regime is similar to the synchronous state in the KS model~\eqref{eq:ks}. Calculations of the phase diffusion constant of the mean field reveal that it decays more strongly with the ensemble size: $D\sim N^{-2.48}$, see Fig.~\ref{fig:3}(b).

A regime where some oscillators are locked by the mean field while others are not is not the only type of synchrony in large ensembles. There exists so-called partial synchronization~\cite{vanVreeswijk-96,Mohanty-Politi-06,Rosenblum-Pikovsky-07,Pikovsky-Rosenblum-09} where no one unit is locked, and nevertheless there is a collective oscillating mode; see \cite{Temirbayev_etal-12,temirbayev2013autonomous} for an experimental observation in a set of electronic generators. As an example of such a type of synchrony, we use a globally nonlinear coupled system of SL oscillators~\eqref{eq:sl1}. For this model, the scaling of the diffusion constant with the system size is nearly the same as for the KS model above: $D\sim N^{-2}$, see Fig.~\ref{fig:3}(a) (best fits for different values of the distribution width of natural frequencies are in the range $(1.92,2.03)$.)

Finally, we explored coherence properties of the collective mode for coupled chaotic oscillators. We took an ensemble of Roessler oscillators~\eqref{eq:roes}. Each oscillator possesses a funnel attractor with ill-defined phase; however, the global mode is nearly periodic with well-defined phase, as has been shown in~\cite{Pikovsky-Rosenblum-Kurths-96,Pikovsky-Rosenblum-24}. Diffusion constant in this case scales like $D\sim N^{-1}$, as shown in Fig.~\ref{fig:3}(c). 

We now explore why, in different cases, the diffusion constant of the global mode shows different dependence on the system size. Let us introduce the instantaneous frequency of the global mode (``global frequency'') defined as $\Omega(t)=\dot\Theta(t)$. Because $\Theta$ is defined via mean fields, the process $\Omega(t)$ is also some kind of a mean field. For example, for the KS model \eqref{eq:ks},\eqref{eq:ksop}, one has $\Omega=(NR)^{-1}\sum_k \cos(\varphi_k-\Theta)(\omega_k+\varepsilon R\sum(\Theta-\varphi_k-\alpha))$. We have found that in all considered examples the variance of the global frequency scales as $\text{var}(\Omega)\sim N^{-1}$. However, it is not the variance of the stationary process $\Omega(t)$ that determines the diffusion constant, but, according to the Green-Kubo formula, the integral of the autocorrelation function (or, in spectral representation, the level of the continuous component of the spectrum at zero frequency). For the decay of the autocorrelation function, one needs the deterministic ensemble to be chaotic. We have characterized this chaoticity by means of the largest Lyapunov exponent $\lambda$. For coupled chaotic Roessler systems, this exponent practically does not depend on the system size $N$, which explains why the diffusion constant in this case scales like $D\sim N^{-1}$, following the scaling of $\text{var}(\Omega)$. In all other cases, the Lyapunov exponent tends to zero in the thermodynamic limit, as has been observed also for the Kuramoto model~\cite{Popovych_etal-05,Carlu-Ginelli-Polit-18}. For the KS model \eqref{eq:ks} and for nonlinearly coupled SL oscillators \eqref{eq:sl1} we have found  $\lambda\sim N^{-1}$, while for the ML systems \eqref{eq:mldet2} we have found $\lambda\sim N^{-0.56}$. This qualitatively explains why the diffusion constant decays strongly in these situations: in addition to a ``standard'' $N^{-1}$ law for the variance of the global frequency, the chaotic component of its variations also vanishes in the thermodynamic limit. We stress here that the largest Lyapunov exponent is only indirectly related to the decay of correlations of the global frequency; in particular, the next example shows that in a chaotic ensemble, the diffusion constant can vanish identically.

\textit{Symmetry transition and quenched diffusion in the Kuramoto model.}
Above we started with the KS model, and not with the ``standard'' Kuramoto model (i.e., Eq.~\eqref{eq:ks} with $\alpha=0$). The reason is that the latter demonstrates rather nontrivial properties of coherence.
Next, we show that the Kuramoto model with a regular sampling of frequencies possesses a symmetric state in which phase diffusion is quenched, and this state is stable over a certain range of parameters. This symmetric state has been mentioned by Daido~\cite{daido1987scaling}, without stability considerations, as a possible regime for corresponding symmetric initial conditions. Stability properties have been first discussed in \cite{Popovych_etal-05,maistrenko2005desynchronization} and explored at small couplings in~\cite{chiba2009stability}. 

For definiteness, we assume the number of units $N=2M$ to be even; for an odd $N$, a simple reformulation of the theory below is needed. We write the Kuramoto system as
\begin{equation}
\begin{gathered}
\dot\vp_k=\w_k+\e(S\cos\vp_k-C\sin\vp_k),\\ C=\frac{1}{N}\sum_{j=1}^N\cos\vp_j,\quad
S=\frac{1}{N}\sum_{j=1}^N\sin\vp_j
\end{gathered}
\label{eq:k1}
\end{equation}
Furthermore, we assume that the frequencies are pairwise symmetric $\w_k=-\w_{-k}$. We introduce new variables
\[
u_m=\frac{\vp_m-\vp_{-m}}{2},\; v_m=\frac{\vp_m+\vp_{-m}}{2},\; m=1,\ldots,M\;.
\]
In this variables the Kuramoto model \eqref{eq:k1} reads
\[
\begin{aligned}
\dot u_m&=\w_m-\e\sin u_m(C\cos v_m+S\sin v_m)\;,\\
\dot v_m&=\e\cos u_m(S\cos v_m-C\sin v_m)\;,\\
&C=M^{-1}\sum_{j=1}^M \cos u_j\cos v_j,\; S=M^{-1}\sum_{j=1}^M \cos u_j\sin v_j\;.
\end{aligned}
\]
One can easily check that $v_1=\ldots=v_M=V=const$ is the invariant manifold of this system of equations, for arbitrary $V$. The arbitrariness of $V$ reflects the invariance of the Kuramoto model with respect to a shift of all the phases. The complex mean field is represented through the introduced quantities as $Z(t)=R(t)e^{i\Theta(t)}=C(t)+iS(t)$ and the instantaneous global frequency is
$\W(t)=\dot\Theta=(\dot S C-\dot C S)/(S^2+C^2)$. On the other hand, on the symmetric manifold
\[
C=Q(t)\cos V,\; S=Q(t)\sin V,\; Q(t)=M^{-1}\sum_{j=1}^M \cos u_j\;.
\]
Substituting this in the instantaneous frequency of the global mode, we obtain that on the symmetric manifold $\dot\W=0$. This means that the variations of the global frequency vanish, and the diffusion is quenched.
On the symmetric manifold, only the variables $u_m$ vary, and in general their dynamics is chaotic. Existence of a symmetric manifold corresponds to the well-known  effect of complete 
synchronization of chaotic systems~\cite{Fujisaka-Yamada-83,Pikovsky-84b,Pikovsky-Rosenblum-Kurths-01}. 

According to the general theory of complete synchronization~\cite{Pikovsky-Rosenblum-Kurths-01, Pikovsky-Politi-16}, stability of the chaotic symmetric state is determined by the largest transversal Lyapunov exponent on the symmetric manifold. To calculate it, we assume a weak violation of symmetry $v_m=V+w_m$, which, after linearization in $w_m$, leads to a system (where $u_m(t)$ is a chaotic trajectory on the symmetric manifold)
\begin{align}
\dot u_m&=\w_m-\e\sin u_m \frac{1}{M}\sum_{j=1}^M\cos u_j\;,
\label{eq:k5-1}\\
\dot w_m&=\e \cos u_m\frac{1}{M}\sum_{j=1}^M \cos u_m(w_j-w_m)\;.
\label{eq:k5-2}
\end{align}
One can see from Eq.~\eqref{eq:k5-2} that there is always a uniform solution $w_1=\ldots=v_M$ with zero Lyapunov exponent; this solution corresponds to a shift of the constant $V$ and is not transversal. To calculate the transversal Lyapunov exponent, we thus set $\sum_j w_j=0$. 

\begin{figure}
\centering
\includegraphics[width=0.9\columnwidth]{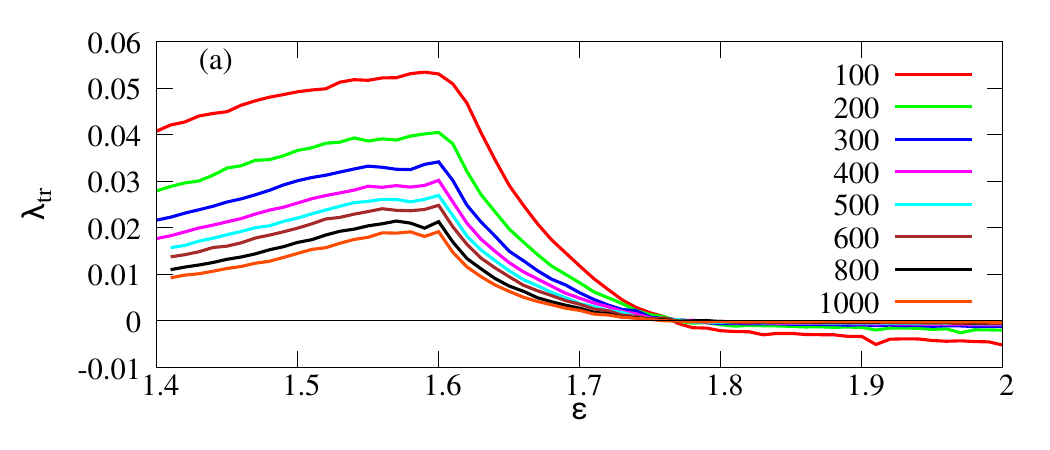}\\[-1em]
\includegraphics[width=0.9\columnwidth]{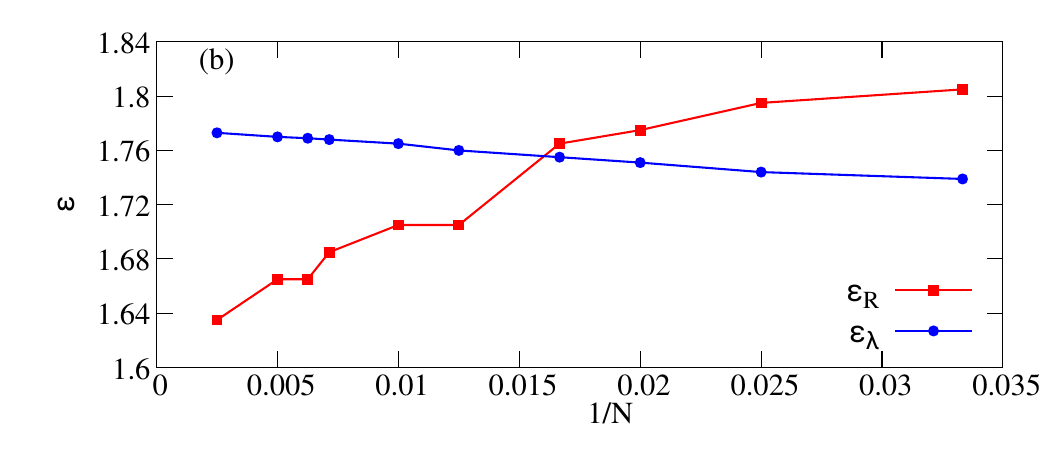}
\caption{Panel (a): The transversal LE vs coupling parameter $\e$ for several ensemble sizes $N$. Panel (b): critical value of coupling $\e$ at which the transversal LE changes sign and the symmetric manifold becomes stable (blue circles). Red squares: critical value of coupling at which a well-defined global mode with $R_{min}>0.1$ appears.}
\label{fig:k1}
\end{figure}

We have calculated the transversal LE $\lambda_{tr}=\lim_{t\to\infty}t^{-1}\ln ||w(t)||/||w(0)||$ for the Kuramoto model with regular sampled Gaussian distribution of frequencies, for a range of system sizes. The results are presented in Fig.~\ref{fig:k1}. The values of the transversal LEs vary significantly with the system size $N$, but the transition point $\e_\lambda$ at which this exponent becomes negative only weakly depends on $N$, as is clear from panel (b), where the value of the coupling constant $\e_\lambda$ at which the transversal Lyapunov exponent changes sign, is presented. In this panel, we also show the value of the coupling $\e_R$, starting from which the minimal value of the order parameter $R$ over a long run becomes larger than $0.1$, which we use as an indicator for the existence of a global mode with a well-defined phase. One can see that the curves cross at $N_s\approx 60$. For smaller system sizes, the symmetric state becomes stable already in the state where the phase is ill-defined; thus, here, no regime with phase diffusion is observed. For $N>N_s$, there is a range of coupling constants $\e_R<\e<\e_{\lambda}$ where the phase is well-defined and diffuses. With an increase of coupling strength, for $\e>\e_\lambda$, a symmetric state occurs with vanishing diffusion. 

\begin{figure}
\centering
\includegraphics[width=\columnwidth]{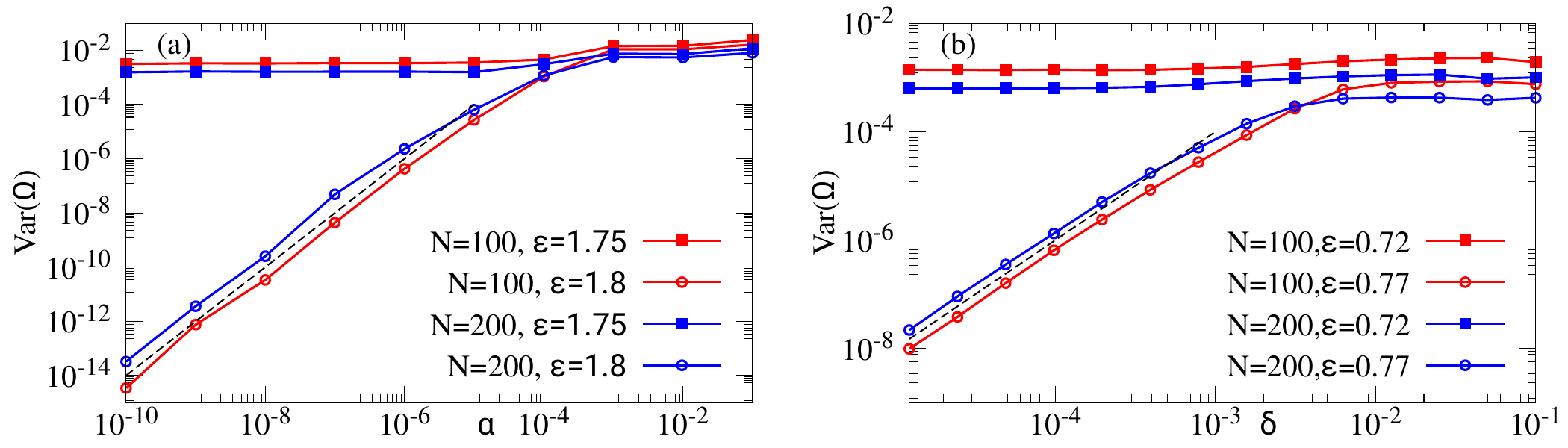}
\caption{Panel (a): The variance of the global frequency vs. phase shift $\alpha$ in the KS system; for two values of coupling strength below and above the symmetry synchronization transition. Panel (b): The variance of the global frequency vs asymmetry parameter $\delta$ of the Beta-distribution of frequencies in the Kuramoto model.}
\label{fig:k2}
\end{figure}

The symmetric state with vanishing global phase dynamics is destroyed if the symmetry of system~\eqref{eq:k1} is broken. We illustrate this in Fig.~\ref{fig:k2} with two types of symmetry breaking.
In the first example (panel (a)), we take the same symmetric natural frequencies as above, but insert a small phase shift $\alpha$ to the coupling terms, making the KS system out of the Kuramoto system. In the second example (panel (b)), we illustrate the situation where the coupling is symmetric, but the set of natural frequencies is slightly skewed. Namely, we sample these frequencies regularly from the beta distribution $B(3+\delta,3-\delta)$ with a small skewness parameter $\delta$. In both cases, the variance of the global frequency grows as the second power of the symmetry-breaking parameter.

\textit{Conclusion.} In this Letter, we explored finite-size induced coherence properties of populations of coupled oscillators. The measure of coherence is the diffusion constant of the global mode's phase. The results  provide a new, richer scenario, complementing previous studies of the amplitude fluctuations and of phase diffusion in an ensemble of noisy units. First, the global phase is ill-defined prior to and in the vicinity of the synchronization transition; we used the condition that the global amplitude is separated from zero to define a region where calculation of the phase diffusion makes sense. Second, in this domain we always observed normal diffusion at which the variance of the phase differences grows proportional to the time lag. Third, we found that if the interacting units are chaotic, then the diffusion scales $D\sim N^{-1}$ with the same exponent as for coupled noisy units. The reason is that the effective disorder in the dynamics does not disappear in the thermodynamic limit, and the scaling is consistent with central-limit-theorem-like arguments. Fourth, for coupled periodic oscillators, we have found much stronger decay of the diffusion constant with the system size, with powers $\approx 2$ for the KS  model and coupled SL oscillators, and $\approx 2.5$ for coupled ML neurons. At the moment, it is not clear if there are different universality classes; one needs to consider more examples. The strong decay of the diffusion constant can be attributed to weak system-size-dependent chaos in these models,
characterized by the largest Lyapunov exponent that vanishes in the thermodynamic limit, which yields additional smallness to the usual $\sim N^{-1}$ decays of mean field fluctuations.  In addition to the general features above, we have shown that the standard Kuramoto model with a symmetric set of natural frequencies exhibits, at sufficiently strong couplings, a symmetric state in which the global phase remains constant in time. We expect this property to hold in other symmetric systems, such as coupled SL oscillators.

\begin{acknowledgments}
\textit{Acknowledgments.}
S.I. acknowledges support from the MUR PRIN2022 project ``Breakdown of ergodicity in classical and quantum many-body
systems'' (BECQuMB) Grant No. 20222BHC9Z.\\
F.B. acknowledges support and hosting from the University of Perpignan via Domitia (France).
\end{acknowledgments}

\appendix
\section{Appendix: Dynamical Models}
Here we describe dynamical models of globally coupled oscillators used in the main text.

\textit{Coupled Roessler systems.}
This model has been studied in Refs.~\cite{Pikovsky-Rosenblum-Kurths-96,Pikovsky-Rosenblum-24}, it reads
\begin{equation}
\begin{aligned}
\dot x_k&=-y_k-z_k+\frac{\e}{N}\sum_j x_j\;,\\
\dot y_k&=x_k+ay_k\;,\\
\dot z_k&=b+(x_k-c)z_k\;.
\end{aligned}
\label{eq:roes}
\end{equation}
We use parameters $a=0.25$, $b=0.4$, $c=8.5$, at which each R\"ossler system has a funnel-type attractor, i.e., the oscillators have an ill-defined phase.
For this system, we use the observables 
\[
X(t)=\frac{1}{N}\sum_{k} x_k(t)\;,\qquad Y(t)=\frac{1}{N}\sum_{k} y_k(t)\;.
\]
The two-dimensional embedding is ($Y(t),X(t)+aY(t)$) because $\dot Y=X+aY$. These observables are normalized as $Y\to\frac{Y-\av{Y}}{STD(Y)}$ and the same for $X+aY$, to obtain a maximally circular embedding. Then the phase of the mean field is calculated as $\arctan \frac{Y}{X+aY}$. The diffusion constant vs $N$ is presented in Fig.~\ref{fig:3}(c).

\textit{Coupled Stuart-Landau (SL) oscillators.} This model has been studied in Ref.~\cite{Rosenblum-Pikovsky-15}. The equations read
\begin{equation}
\begin{gathered}
\dot a_k=(1+i\omega_k)a_k-|a_k|^2a_k+(\e_1+i\e_2)A-\eta|A|^2A,\\
A=\frac{1}{N}\sum_k a_k\;.
\end{gathered}
\label{eq:sl1}
\end{equation}
Parameters are $\e_1=\e_2=3,\quad \eta=10$. Note that the nonlinear mean field means many-body coupling (quadruplets), because
$A|A|^2=N^{-3}\sum_{jlm} a_j a_l a_m^*$. In Fig.~\ref{fig:3}(a), we show diffusion constants vs $N$ for a uniform distribution of natural frequencies in an interval $(-q,q)$, for several values of $q$.

\textit{Coupled Morris-Lecar neurons.}
The ML model is a two-dimensional system describing spiking neurons. It is formulated in terms of variables $(V,n)$, where $V$ is voltage and $0\leq n\leq 1$ is the portion of open  
potassium channels (the number of sodium channels $m$ is taken from its equilibrium value $m_\infty$ at a given voltage.) We adopt, following \cite{tanaka2014dynamical}, the following model of $N$ synaptically coupled ML systems ($k=1,\ldots,N$; $\av{V}=N^{-1}\sum_k V_k$):
\begin{equation}
\begin{aligned}
\frac{d V_k}{dt}&=\frac{1}{C}\big(\e(\av{V}-V_k)+I_{ext,k}-g_{Ca}m_{\infty,k} (V_k-V_{Ca})-\\
&-g_L(V_k-V_L)-g_K n_k (V_k-V_K)\big),\\
\frac{dn_k}{dt}&=(1-n_k)\alpha(V_k)-n_k\beta(V_k).\\
&m_{\infty,k}=\frac{1}{2}\left(1+\tanh(\frac{V_k-V_a}{V_b})\right),\\
&\alpha(V)=\phi\frac{\cosh (\xi/2)}{1+e^{-2\xi}},\quad
\beta(V)=\phi\frac{\cosh (\xi/2)}{1+e^{2\xi}},\\ & \xi=\frac{V-V_c}{V_d}.
\end{aligned}
\label{eq:mldet2}
\end{equation}
 Inhomogeneity in the population is introduced by uniformly distributed external currents $100\leq I_{ext,k}\leq 130$. Other parameters are $
C=20$; $V_K=-84$; $V_L=-60$; $V_{Ca}=120$; $g_K=8$;  $g_L=2$;  $g_{Ca}=4.4$;  $V_c=2$;  
$V_d=30$;     $\phi=0.04$.
A synchronized state where approximately half of the oscillators are locked by the mean field is achieved at $\e=0.28$, for this coupling, the diffusion constants are shown in Fig~\ref{fig:3}(b).

%

\end{document}